\documentclass[aps,prl,twocolumn,superscriptaddress, floatfix]{revtex4-2}%
\usepackage{times,mathptmx}

\usepackage{float}  

\usepackage{graphicx}
\usepackage{graphics}
\usepackage[caption=false]{subfig}
\usepackage{color}
\usepackage[english]{babel}
\usepackage[utf8]{inputenc}
\usepackage{ae}
\usepackage{latexsym}
\usepackage{dcolumn}
\usepackage{bm}
\usepackage[table]{xcolor}
\usepackage{array}
\usepackage{url}
\usepackage{epsf}
\usepackage{epsfig}
\usepackage{pstricks,pst-grad}
\usepackage[pdftex,colorlinks]{hyperref}
\hypersetup{
    colorlinks=true,
    urlcolor=blue,
    }
\usepackage{amsmath}
\usepackage{amssymb}
\usepackage{ragged2e}
\usepackage{comment}

\begin{document}

\title[]
{Anomalous Long-range Hard-wall Repulsion between Polymers in Solvent Mixtures and Its Implication for Biomolecular Condensates}

\author{Luofu Liu}
\affiliation{Department of Chemical and Biomolecular Engineering, University of California Berkeley, Berkeley, California 94720, United States}
\affiliation{Materials Sciences Division, Lawrence Berkeley National Lab, Berkeley, California 94720, United States}

\author{Rui Wang}
\email {ruiwang325@berkeley.edu}\affiliation{Department of Chemical and Biomolecular Engineering, University of California Berkeley, Berkeley, California 94720, United States}
\affiliation{Materials Sciences Division, Lawrence Berkeley National Lab, Berkeley, California 94720, United States}


\begin{abstract}

The system of polymers in solvent mixtures is a widely-used model to represent biomolecular condensates in intracellular environments. Here, we apply a variational theory to control the center-of-mass of two polymers and perform the first quantification of their interactions in solvent mixtures. Even both solvent and cosolvent are good to the polymer, we demonstrate that strong polymer-cosolvent affinity induces the formation of a single-chain condensate. Even though all the molecular interactions are soft, the potential of mean force between two condensates exhibits an anomalous feature of long-range hard-wall repulsion, which cannot be categorized into any existing types of inter-chain interactions. This repulsion is enhanced as either the affinity or the bulk cosolvent fraction increases. The underlying mechanism is cosolvent regulation manifested as a discontinuous local condensation of cosolvent. The hard-wall repulsion provides a kinetic barrier to prevent coalescence of condensates and hence highlights the intrinsic role of proteins as a cosolvent in stabilizing biomolecular condensates.              

\end{abstract}
\maketitle

The conformation and interaction of polymers in solvents not only govern the thermodynamic and dynamic properties of polymer solutions \cite{rubinstein2003polymer, des1990polymers, fujita2012polymer, fredrickson2006}, but are also fundamentally important to study complex soft-matter systems, such as self-assembled micelles \cite{Mai2012a}, polymer-grafted nanoparticles \cite{Hore2019}, coacervates \cite{Rumyantsev2021}, and protein aggregates \cite{Ross2004}. While the behaviors of polymers in a single solvent have been extensively studied, many practical applications involve multiple solvents to regulate the solubility and diversify the functionality. The addition of cosolvents greatly enriches the conformational and phase behaviors, even leading to counterintuitive phenomena \cite{Altena1982, Shultz1955, Termonia1999, Wolf1978}. One telling example is the ``cononsolvency effect": polymer chain collapses in a mixture of two good solvents, seemingly defying the law of solubility \cite{Wolf1978, Mukherji2017, Bharadwaj2022}. 

Solvent mixtures also widely exist in biological systems \cite{Sommer2022, Haugk2024, Pappu2023}. From a physical chemistry standpoint, the intracellular environment is essentially a system of biomacromolecules dissolved in a complex solvent mixture comprising water, salts and metabolites \cite{Pappu2023}. The interplay between these multiple components is critical to maintain the structure and functionality of cells. Particularly, liquid-liquid phase separation (LLPS) drives the formation of condensates which compartmentalize cells to organize and regulate biochemical reactions \cite{Banani2017, Kilgore2022, Julicher2024, Pappu2023}. Despite many efforts, it remains unclear how these condensates are stabilized at the stage of finite-size droplets against Ostwald ripening and coalescence \cite{Mittag2022}. Depending on specific intracellular environments, different mechanisms are hypothesized including physical barriers \cite{Feric2013, Lafontaine2021, Kim2021}, electrostatic repulsions \cite{Welsh2022}, chemical reactions \cite{Wurtz2018}, and active processes \cite{Soding2020, Cates2024, Weber2019}. Many nuclear and cytoplasmic condensates are rich in RNA and RNA-binding proteins, where RNA seeds the nucleation of the condensates and proteins can be viewed as cosolvents \cite{Roden2021, Garcia-JoveNavarro2019}. Recent experiments by Folkmann et al. observed that an intrinsically disordered protein, MEG-3, significantly slows the coarsening of P-granules (a typical RNA-protein complex) \cite{Folkmann2021}. It is explained by the adsorption of MEG-3 at the condensate surface as Pickering agent to reduce the surface tension and inhibit the coalescence \cite{Folkmann2021, Pickering1907, Chevalier2013}. Similar surface enrichment of proteins has also been found in other biomolecular condensates \cite{Visser2024, Linsenmeier2023, Adame-Arana2023, Kelley2021, Abeysinghe2024, Feric2022, Folkmann2021}. However, such ``surfactant-like" explanation contradicts the observation from fluorescence microscopy that proteins can also distribute in the interior of the condensates\cite{Folkmann2021}. 

Pioneered by Pappu and Brangwynne, there is a growing interest in studying intracellular organizations and phase transitions from the polymer physics perspective \cite{Brangwynne2015, Pappu2023, Shin2017, Lee2013, Hyman2014, Zhou2024}. Flory-Huggins (FH) theory and its modifications have been widely used to model the phase equilibrium and dynamics of biomolecular condensates \cite{Nott2015, Berry2015,Saha2016}. Lee et al successfully explained the assembly of P-granules in the mixture of protein MEX-5 and cytoplasmic constituents \cite{Lee2013}. Combining FH theory and molecular simulations, Sommer et al. studied the condensation of nucleosomes in the presence of water and proteins \cite{Sommer2022, Haugk2024}. By accounting for the preferential interaction between nucleosomes and proteins, they proposed a mechanism of polymer-assisted condensation. While the above theories successfully explained why condensates form from individually water-soluble components, an important aspect is still puzzling: why are condensates stabilized to finite sizes without coalescence? An important factor that FH-based theories fails to capture is the microscopic heterogeneity: component densities spatially vary across the condensate interface. This is critical to understand the structure of the condensates and their interactions. To our knowledge, the interactions between two biomolecular condensates, and even a more fundamental problem of the interactions between two polymers in a solvent mixture, has not been studied yet.   

In our earlier work, we developed a variational theory to control the center of mass (c.m.) of polymers \cite{Liu2024a}. Here, we modify it to model polymers in a solvent mixture. Even both solvent and cosolvent are good to the polymer, we demonstrate that a single-chain condensate can form due to strong polymer-cosolvent affinity. We also investigate the interactions between two condensates via quantifying the potential of mean force (PMF), which surprisingly exhibits a long-range hard-wall repulsion as a result of cosolvent regulation.

We consider a system of two identical polymer chains (P1 and P2) immersed in $n_{\rm S}$ solvent and $n_{\rm C}$ cosolvent molecules. The system is connected to a reservoir of binary mixture of solvents to maintain fixed chemical potentials $\mu_{\rm S}$ and $\mu_{\rm C}$ of solvent and cosolvent, respectively. The solvent is treated as a small molecule and the cosolvent is treated as an oligomeric chain. This enables our system to model biomolecular condensates, where the polymer represents RNA which seeds the nucleation of the condensates \cite{Garcia-JoveNavarro2019} and the oligomeric cosolvent represents RNA-binding protein. Both the polymer and the cosolvent are described by the continuous Gaussian-chain containing $N$ and $N_{\rm C}$ Kuhn segments, respectively, with Kuhn length $b$. The volumes of a Kuhn segment and a solvent are assumed to be the same $v$. To study the interaction between two polymers, we fix their c.m. at $\bm{\xi}_1$ and $\bm{\xi}_2$ respectively with the c.m. separation distance $L = |\bm{\xi}_2-\bm{\xi}_1|$. The semi-canonical partition function is
\begin{equation}
\begin{aligned}
    &Z = \sum_{n_{\rm S}, n_{\rm C} =0}^{\infty}\frac{e^{\beta\mu_{\rm S}n_{\rm S}} e^{\beta\mu_{\rm C}n_{\rm C}}}{n_{\rm S}!n_{\rm C}!v^{2N +n_{\rm S} +n_{\rm C}N_{\rm C}}} \\
    &\prod_{\alpha=1}^2 \int\hat{{\rm D}}\{\bm{R}_{\alpha}\} \prod_{\gamma=1}^{n_{\rm S}} \int {\rm d} {\bm{r}_\gamma} \prod_{\kappa=1}^{n_{\rm C}} \int\hat{{\rm D}}\{\bm{R}_{\kappa}\} \exp(-\beta H)\\
    &\prod_{\bm{r}} \delta \left[1 - \sum_{j={\rm P1}, {\rm P2}, {\rm S}, {\rm C}} \hat{\phi}_j \right] \prod_{\alpha=1}^2 \delta \left[\frac{1}{N}\int_0^N {\rm d}s \bm{R}_\alpha -\bm{\xi}_\alpha \right]
\end{aligned}
\label{eq:particle_partition_function}
\end{equation}
where $\int \hat{\rm D}\{\bm{R}\}$ denotes the functional integration over all possible chain configurations weighted by Gaussian statistics. $\beta H = v^{-1}\sum_{j,k}\chi_{jk}\int{\rm d}\bm{r}\hat{\phi}_j\hat{\phi}_k$ ($j,k={\rm P}, {\rm S}, {\rm C}$) includes interactions among the polymers, solvents, and cosolvents in terms of the Flory-Huggins $\chi$ parameters. $\hat{\phi}_{\rm P} = \hat{\phi}_{\rm P1} + \hat{\phi}_{\rm P2}$ is the total instantaneous volume fraction of polymer. The first $\delta$-functional accounts for the incompressibility and the second enforces the c.m. constraint on the two polymers.

We follow the variational approach developed in our previous work \cite{Liu2024a} by decomposing $Z$ as $Z=Z_0Z_{F1}Z_{F2}$, where $Z_0$ is the contribution irrelevant to fixing the c.m. and $Z_{F\alpha}$ ($\alpha=1,2$) comes from the constraint of c.m. of chain $\alpha$. $Z_{F\alpha}$ is given by

\begin{equation}
    Z_{F\alpha} = \int {\rm d} \bm{F}_\alpha \exp(-L_{{F}\alpha})
\end{equation}
where $\bm{F}_\alpha$ is the force field conjugate to the deviation of c.m. of Polymer $\alpha$ with the ``action" $L_{F\alpha} = -\ln Q_\alpha$, where $Q_\alpha$ is the single-chain partition function in the auxiliary fields $W_{\rm{P}\alpha}$ (conjugate to polymer density) and $\bm{F}_\alpha$:
\begin{equation}
\label{eq:single_chain_partition_function}
\begin{aligned}
    &Q_\alpha = \frac{1}{v^{N}}\int {\rm \hat{D}} \{{\bm{R}_\alpha}\} \\
    &\exp \bigg\{ -\int_0^N {\rm d}s \bigg[ iW_{{\rm P}\alpha}(\bm{R}_\alpha(s))
    - i\frac{\bm{F}_\alpha}{N} \cdot (\bm{R}_\alpha(s) - \bm{\xi}_\alpha)   \bigg]  \bigg\}
\end{aligned}
\end{equation}
To account for the fluctuations of c.m., we perform a non-perturbative variational approach using the Gibbs-Bogoliubov-Feynman bound \cite{kardar2007} to evaluate $Z_{F\alpha}$:

\begin{equation}\label{eq:Feynman_bound}
    Z_{F\alpha} \approx Z_{\rm ref,\alpha} \exp[-\langle L_{F\alpha} - L_{\rm ref,\alpha} \rangle_{\rm ref}]
\end{equation}
where the average $\langle ... \rangle_{\rm ref}$ is taken in the reference ensemble with action $L_{\rm ref,\alpha}$. $Z_{\rm ref,\alpha}=\int {\rm d} \bm{F}_\alpha \exp(-L_{\rm ref,\alpha})$ is the normalization factor. For mathematical convenience, $L_{\rm ref,\alpha}$ is taken to be a $2n$-power modified Gaussian
\begin{equation}
    e^{-L_{\rm ref,\alpha}} = \prod_{\kappa = x,y,z} (F_{\alpha, \kappa} + if_{\alpha,\kappa})^{2n} \exp \left[ - \frac{(F_{\alpha,\kappa} + if_{\alpha,\kappa})^2}{2A_{\alpha, \kappa}} \right]
\end{equation}
where the average force $\bm{f}_\alpha$ and coefficients $A_{\alpha,\kappa}$ are taken to be the variational parameters. This construction of $L_{\rm ref,\alpha}$ assumes that c.m. fluctuates independently in the three Cartesian directions. A general $2n$-power modified Gaussian is adopted here instead of a standard ($0$th-power) Gaussian to reinforce the confinement of the c.m., which is necessary to confine a polymer in coil state \cite{Liu2024a}.

Minimizing $Z$ with respect to $\bm{f}_\alpha$, $A_{\alpha,\kappa}$ and taking saddle-point approximation for other fields, we obtained the following constrained self-consistent equations (see Sec. I in the Supplementary Materials for the detail derivation):
\begin{subequations}\label{eq:scf}
\begin{align}
    &w_{\rm P\alpha}(\bm{r})  = \chi_{\rm PS}\phi_{\rm S}(\bm{r})  + \chi_{\rm PC}\phi_{\rm C}(\bm{r}) +\gamma(\bm{r}) \\
    &w_{\rm S}(\bm{r})  = \chi_{\rm PS}\phi_{\rm P}(\bm{r})  + \chi_{\rm SC}\phi_{\rm C}(\bm{r}) +\gamma(\bm{r}) \\
    &w_{\rm C}(\bm{r})  = \chi_{\rm PC}\phi_{\rm P}(\bm{r})  + \chi_{\rm SC}\phi_{\rm S}(\bm{r}) +\gamma(\bm{r}) \\
    &\phi_{\rm P\alpha}(\bm{r}) = Q_\alpha^{-1}\int {\rm d}s q_\alpha(\bm{r},N-s)q_\alpha(\bm{r},s) \\
    & \phi_{\rm S}(\bm{r}) = \exp[\beta \mu_{\rm s} - w_{\rm s}(\bm{r})]  \\
    &\phi_{\rm C}(\bm{r}) = e^{\beta\mu_{\rm C}}\int {\rm d}s q_{\rm C}(\bm{r},N-s)q_{\rm C}(\bm{r},s) \\
    &  \int {\rm d} \bm{r} (\bm{r} - \bm{\xi}_\alpha) \phi_{\rm P\alpha}(\bm{r}) = \bm{0}  \\
    & A_{\alpha, \kappa} = \lambda / R_{\alpha, \kappa}^2 
\end{align}
\end{subequations}
where $w_j$ and $\gamma$ are the equilibrium value of the fields conjugate to density $\phi_j$ and the incompressibility condition, respectively. The single-chain partition function of the Polymer $\alpha$ is given by $Q_\alpha = v^{-1} \int {\rm d} \bm{r} q_\alpha(\bm{r},N-s)q_\alpha(\bm{r},s)$, where the chain propagator $q_\alpha$ satisfies:
\begin{equation}\label{eq:propagator}
    \frac{\partial q_\alpha}{\partial s} = \frac{b^2}{6}\nabla^2 q_\alpha(\bm{r} ,s)- U_{\rm P\alpha}(\bm{r}) q_\alpha(\bm{r} ,s)
\end{equation}
The total field experienced by Polymer $\alpha$ is
\begin{equation}
    U_{\rm P\alpha}(\bm{r}) = w_{\rm P\alpha}(\bm{r}) - \frac{1}{N} \bm{f}_\alpha\cdot (\bm {r} - \bm{\xi}_\alpha) + \sum_{\kappa = x,y,z}
\frac{\lambda (\kappa-\xi_\kappa)^2}{2N R_{\alpha,\kappa}^2}
\end{equation}
where $R_{\alpha,\kappa}$ is the $\kappa$-th component of the radius of gyration. The prefactor $\lambda$ in the spring constant comes from the power index $n$ of the modified Gaussian. It is calibrated in a single solvent using the criterion $R_\alpha^2 = Nb^2/6$ at $\chi=0.5$ as $N \to \infty$ \cite{Liu2024a}. The cosolvent propagator follows the same equation as Eq. \ref{eq:propagator} with $U_{\rm P\alpha}$ replaced by the field $w_{\rm C}$. The theory enables a simultaneous confinement of two individual polymers. This confinement is achieved by a mean force $\bm{f}_\alpha$ that controls the c.m. position and a harmonic potential with the spring constant $\lambda/(N R_{\alpha,\kappa}^2)$ that counters the c.m. fluctuation.

The theory is general for various combinations of polymers and solvent mixtures. To better represent RNA-protein condensates in water, we assume that solvent and cosolvent are miscible but repulsive to each other ($\chi_{\rm SC}=0.8$), representing proteins with hydrophobic backbones. The bulk cosolvent volume fraction $\phi_{\rm C}^{\rm bulk}$ is chosen outside the binodal of the binary mixture. The quality of solvent is assumed good to the polymer with $\chi_{\rm PS}=0$, consistent with the hydrophilicity of RNA. Furthermore, we assume an attractive interaction between the polymer and cosolvent by setting $\chi_{\rm PC} < 0$ to capture the RNA-protein affinity. We take $N=200$ and $N_{\rm C}=20$ as typical chain lengths of the polymers and the oligomeric cosolvent. The chain stiffness is set to $b^3/v = 1$, leading to a calibrated $\lambda=3.61$ \cite{Liu2024a}. To focus on the cosolvent effect on the condensation, we vary $\chi_{\rm PC}$ and $\phi_{\rm C}^{\rm bulk}$ while maintaining other parameters constant.

\begin{figure}[t] 
\centering
\includegraphics[width=0.48\textwidth]{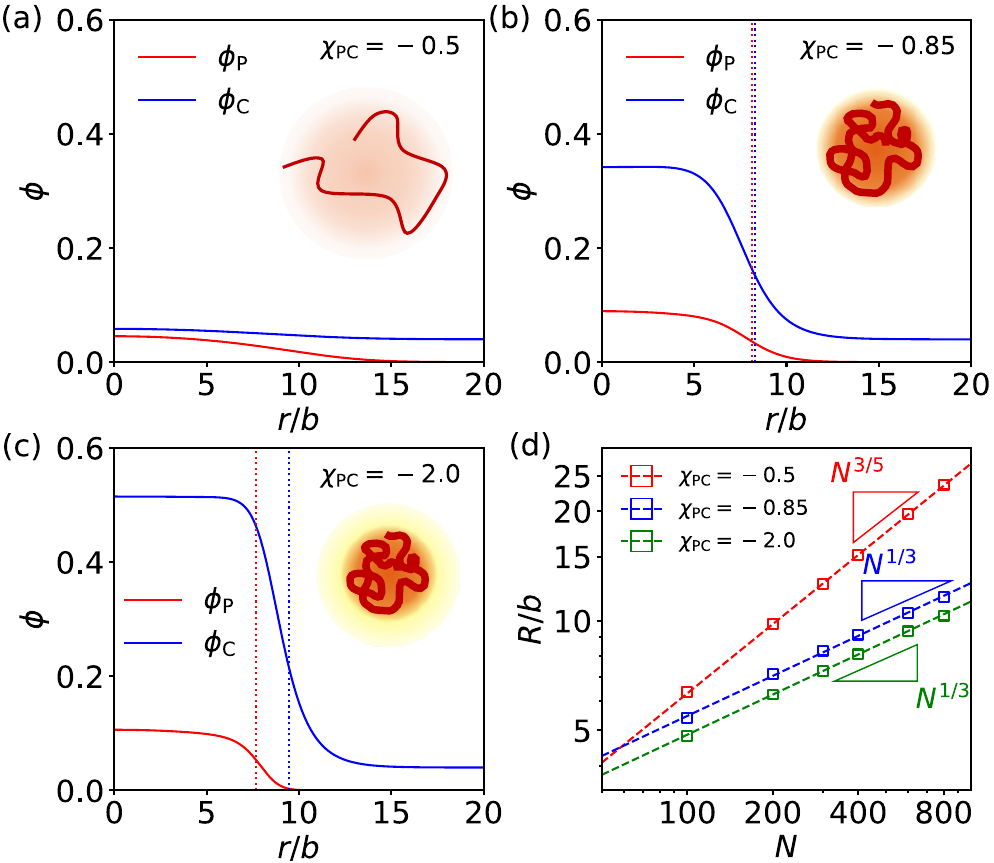}
\caption{The structure of condensate formed by a polymer in solvent-cosolvent mixture. (a)-(c) Density profiles of polymer ($\phi_{\rm P}$) and cosolvent ($\phi_{\rm C}$) for different affinities $\chi_{\rm PC}$. The dotted lines in (b) and (c) denotes the Gibbs dividing interfaces based on the polymer and cosolvent densities, respectively. The insets illustrate the polymer (brown) overlaying on the cosolvent region (yellow). (d) Scaling relationship of the polymer radius of gyration $R$ versus $N$. $\phi_{\rm C}^{\rm bulk}=0.04$.} 
\label{figure:single-chain_profile}
\end{figure}

We first study the conformation of an isolated polymer. Figures \ref{figure:single-chain_profile}a-c plot the density profiles of polymer and cosolvent for different affinities $\chi_{\rm PC}$. As $\chi_{\rm PC}$ becomes more negative, polymer undergoes a conformational transition from a swollen coil ($\chi_{\rm PC}=-0.5$) to a globular condensate ($\chi_{\rm PC}=-0.85$ and $-2.0$). This is confirmed by the change of the scaling of the radius of gyration from $R \sim N^{3/5}$ to  $R \sim N^{1/3}$ as shown in Fig. \ref{figure:single-chain_profile}d. The condensate consists of a core with uniform density and a diffuse interface. Seemingly defying the law of solubility, the polymer can collapse to a condensate even though both the solvent and cosolvent are good to it. The formation of the condensate is energetically driven: a large amount of cosolvents condense to the polymer to enhance their contact.

Compared to polymer globule in a single solvent, the condensate observed in the solvent mixture shows interesting interfacial structure. As the affinity increases from $\chi_{\rm PC}=-0.85$ to $-2.0$, the polymer and the cosolvent interfaces are separated (see Fig. \ref{figure:single-chain_profile}c). The cosolvent interface extends further to the bulk, such that the polymer is enclosed by a layer of excessive cosolvent. This layer replaces the polymer-solvent contact with the more favorable polymer-cosolvent contact. The separation of interfaces predicted by our theory is consistent with the intensity profile of P-granules measured by fluorescence microscopy \cite{Folkmann2021}. Furthermore, we note that the polymer-assisted condensation has also been observed in earlier simulations by Sommer et al \cite{Sommer2022}. However, our work provides the first depiction of the interfacial structure, which plays a key role in understanding the stability of the condensate as we will discuss later. 

\begin{figure}[t] 
\centering
\includegraphics[width=0.48\textwidth]{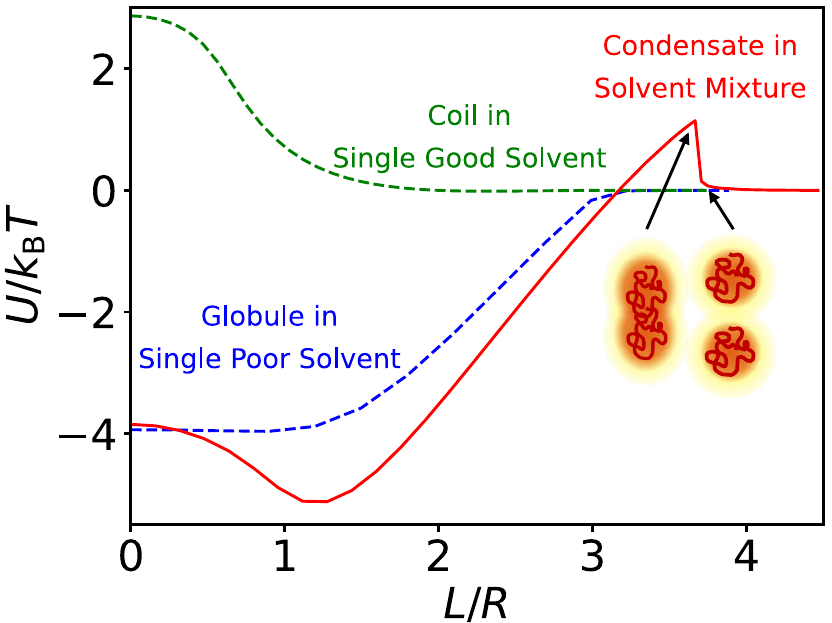}
\caption{Potential of mean force between two polymer-cosolvent condensates in a solvent mixture, in comparison to polymers in a single good/poor solvent. The c.m. separation distance $L$ is rescaled by the polymer radius of gyration $R$. The insets illustrate the two states before and after the discontinuous jump. For solvent mixture, $\chi_{\rm PC}=-1.7$, $\phi_{\rm C}^{\rm bulk}=0.04$. $\chi_{\rm PS}=0.0$ and $1.0$ for single good/poor solvent, respectively.}
\label{figure:PMF_comparison}
\end{figure}

Next, we investigate the interaction between two polymers in the solvent mixture via controlling their c.m. Our theory allows the direct quantification of the PMF $U(L)$, the free energy at separation $L$ excess to two infinitely separated polymers. Figure \ref{figure:PMF_comparison} plots the PMF of two polymer-cosolvent condensates. For comparison, we also provide PMFs of two classical cases: two coil polymers in a single good solvent and two globular polymers in a single poor solvent. The PMF of two condensates is strikingly different from either of the two classical cases: it shows an anomalous long-range hard-wall repulsion as the two condensates is about to contact. The hard-wall repulsion is manifested as a dramatic increase of $U$ in an infinitely short range of $L$. This repulsion is ended by a jump to an attraction between two merged condensates, as indicated by the discontinuity of the slope of $U$ at $L=3.5R$. Counterintuitively, adding a cosolvent can qualitative change the PMF of polymers, which cannot be categorized into any existing types nor be obtained by the superposition of them. It is also surprising that the condensates in the solvent mixture exhibit a hard-wall repulsion, given all components and interactions are soft.

The long-range hard-wall repulsion has great impact on the stability of condensates. Although the association of the two condensates is thermodynamically favorable (indicated by the global minimum in the PMF), the hard-wall repulsion yields an energy barrier which kinetically prevents their coalescence. This provides a possible explanation to the stability of RNA-protein condensates.     

\begin{figure}[t] 
\centering
\includegraphics[width=0.48\textwidth]{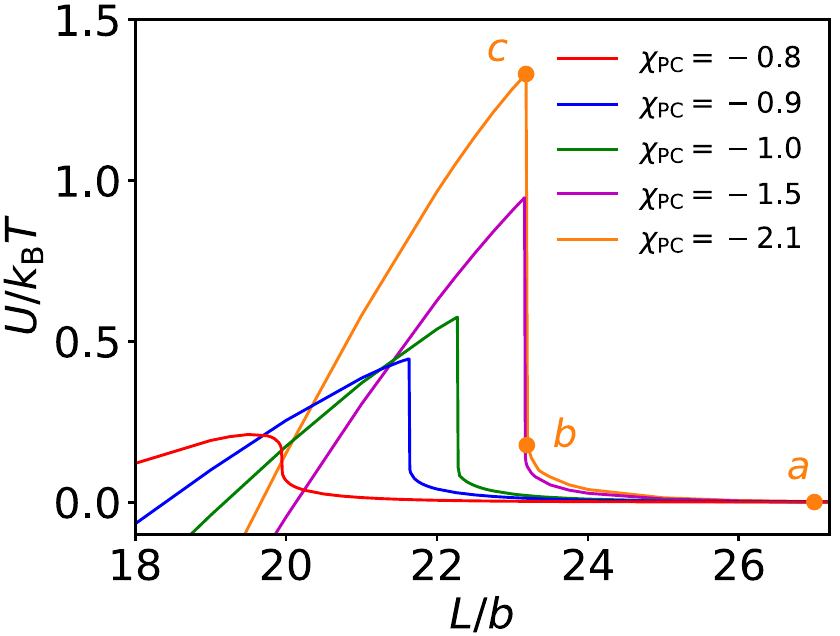}
\caption{Effect of polymer-cosolvent affinity $\chi_{\rm PC}$ on the long-range hard-wall repulsion. $\phi_{\rm C}^{\rm bulk}=0.04$.} 
\label{figure:PMF_change_chiPC}
\end{figure}

Figure \ref{figure:PMF_change_chiPC} plots the effect of polymer-cosolvent affinity on the PMF. The association is continuous at low affinity ($\chi_{\rm PC}=-0.8$). As affinity increases, the transition from repulsion to attraction becomes sharper and turns discontinuous. The hard-wall repulsion also becomes stronger, indicating that the stability of the condensates is enhanced with the affinity. Meanwhile, the energy barrier shifts towards larger $L$, consistent with a thicker cosolvent excess layer outside the polymer, leading to a longer effective range of the repulsion. 

\begin{figure}[t] 
\centering
\includegraphics[width=0.48\textwidth]{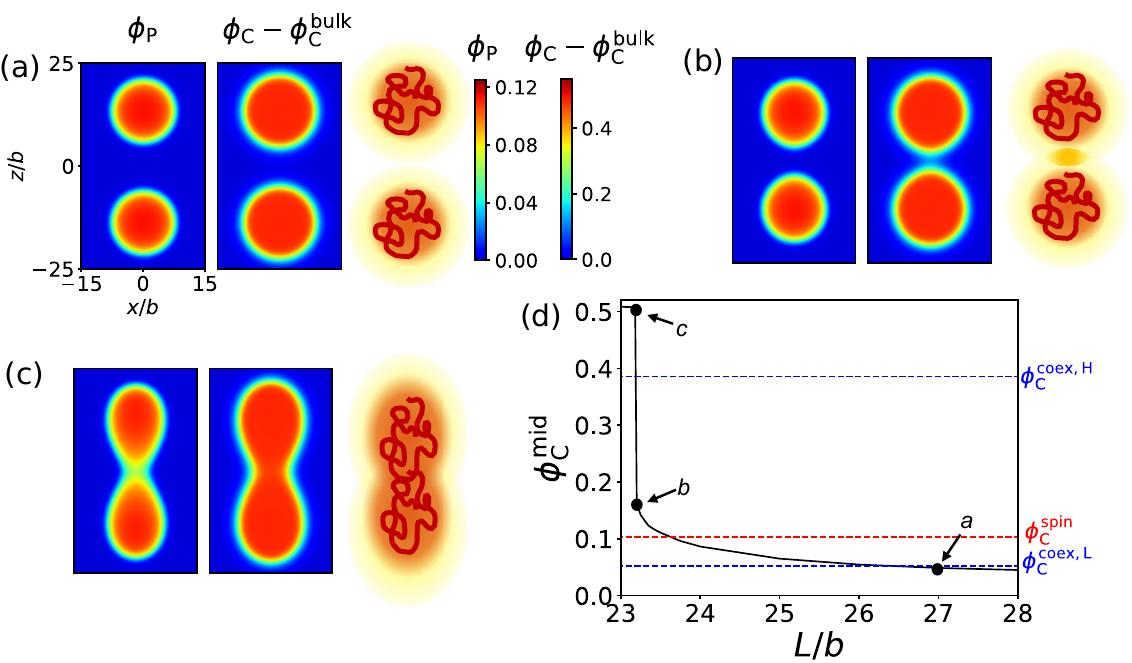}
\caption{The mechanism of hard-wall repulsion as cosolvent regulation. (a)-(c) 2D Density profiles of polymer $\phi_{\rm p}$ and cosolvent (excess to the bulk) $\phi_{\rm C}-\phi_{\rm C}^{\rm bulk}$ plotted in the $x$-$z$ plane for States \textit{a}, \textit{b}, and \textit{c} in Fig. \ref{figure:PMF_change_chiPC}. The c.m. separation distances $L/b$ are respectively $27$, $23.19$ and $23.18$. Schematics of the two interacting condensates are shown on the right. (d) Cosolvent density $\phi_{\rm C}^{\rm mid}$ at the middle point of the center line ($x,y,z=0$) between the two condensates as a function of $L$.} 
\label{figure:schematic_hard_wall}
\end{figure}

The underlying mechanism of the anomalous long-range hard-wall repulsion can be interpreted as ``cosolvent regulation". Figure \ref{figure:schematic_hard_wall}a-c visualize representative states along the PMF (corresponding to States \textit{a}-\textit{c} in Fig. \ref{figure:PMF_change_chiPC}). We also plot cosolvent density $\phi_{\rm C}^{\rm mid}$ at the middle point of the center line ($x,y,z=0$) in Fig. \ref{figure:schematic_hard_wall}d. At State \textit{a}, the two condensates are far away from each other. $\phi_{\rm C}^{\rm mid}$ maintains the value in the bulk mixture which is below the coexistence line of the cosolvent-dilute branch, $\phi_{\rm C}^{\rm coex, L}$. As two condensates approach, cosolvent density starts to overlap, while the two polymers remain isolated due to the protection from the cosolvent excess layer. Meanwhile, $\phi_{\rm C}^{\rm mid}$ increases and exceeds both the coexistence line and the spinodal $\phi_{\rm C}^{\rm spin}$. At State \textit{b}, cosolvent overlapping is pronounced and $\phi_{\rm C}^{\rm mid}$ is deep in the spinodal region. A slight increase of $\phi_{\rm C}^{\rm mid}$ due to a small compression makes the state of the local cosolvent unstable, triggering a discontinuous jump to State \textit{c}. Cosolvent is condensed locally and forms a cosolvent-rich channel connecting polymers. This is confirmed by a dramatic increase of $\phi_{\rm C}^{\rm mid}$ to the value above the coexistence line of the cosolvent-concentrated branch, $\phi_{\rm C}^{\rm coex, H}$. Meanwhile, with the aid of the cosolvent-rich channel, the two polymers deform and merge via neck formation as shown in Fig. \ref{figure:schematic_hard_wall}c. The strong accumulation of cosolvent in a very short range of $L$ leads to dramatic increase of energy and hence a hard-wall repulsion. This can be explained by the energetically-unfavorable local accumulation of cosolvent given the bulk solvent mixture is outside the coexistence line. The molecular picture highlights the vital role of cosolvent in generating the hard-wall repulsion and regulating the stability of the condensates. 

\begin{figure}[t] 
\centering
\includegraphics[width=0.48\textwidth]{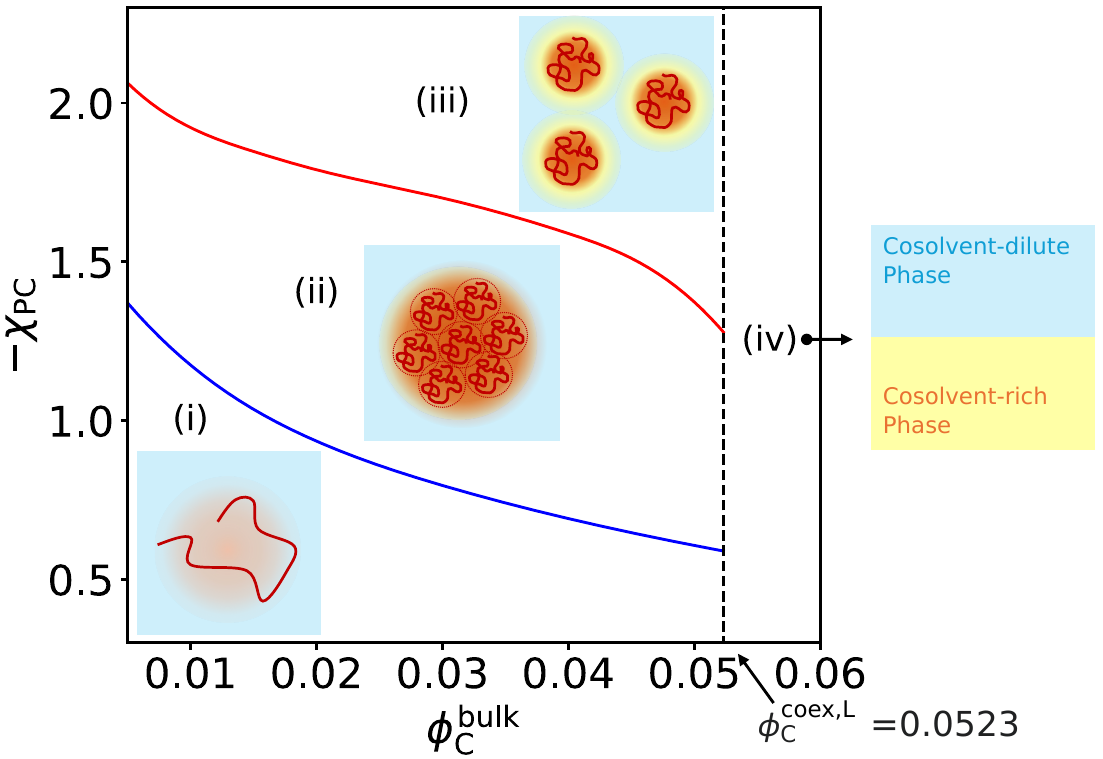}
\caption{State diagram in terms of affinity $-\chi_{\rm PC}$ and cosolvent fraction in the bulk mixture $\phi_{\rm C}^{\rm bulk}$ divided into four regimes: (i) isolated coil, (ii) giant droplet consisting of multiple condensates, (iii) kinetically-stable isolated condensates, and (iv) macrophase separation of solvent and cosolvent. The boundary between (i) and (ii) is located via the steepest change of polymer radius of gyration $R$. The boundary between (ii) and (iii) is determined where the energy barrier in the PMF equals $k_{\rm B}T$.} 
\label{figure:phase_diagram}
\end{figure}

The hard-wall repulsion in the PMF allows us to define a criterion for the stability of an individual condensate. If the energy barrier is comparable to $k_{\rm B}T$, the two-condensate association becomes sufficiently slow, such that condensates are kinetically stable against coalescence. To systematically elucidate the role of cosolvent, Fig. \ref{figure:phase_diagram} plots a state diagram in terms of affinity $-\chi_{\rm PC}$ and bulk cosolvent fraction in the mixture $\phi_{\rm C}^{\rm bulk}$. When both $-\chi_{\rm PC}$ and $\phi_{\rm C}^{\rm bulk}$ are small, there is no sufficient driving force to form a condensate. The polymer behaves as a coil in good solvent (Regime i). In Regime ii, condensate is formed as either $-\chi_{\rm PC}$ or $\phi_{\rm C}^{\rm bulk}$ increases. However, the repulsive energy barrier in the PMF is below $k_{\rm B}T$. Individual condensates are not stable and will coalesce to a giant droplet. In Regime iii, Further increasing $-\chi_{\rm PC}$ and $\phi_{\rm C}^{\rm bulk}$ lifts the repulsive energy barrier above $k_{\rm B}T$. Although coalescence is thermodynamically more favorable due to a larger driving force of condensation, the energy barrier is so high that individual condensates are kinetically stable. The system is thus trapped as a dispersion of isolated single-chain condensates. When $\phi_{\rm C}^{\rm bulk} > \phi_{\rm C}^{\rm coex,L}$ in Regime iv, the solvent and cosolvent are immiscible; macrophase separation occurs.

In this Letter, we modified the variational theory to control the c.m. of two polymers and investigated their interactions in a solvent mixture. We found that polymer-cosolvent affinity can induce the formation of single-chain condensate even though both solvent and cosolvent are good to the polymer. The PMF between two condensates exhibits an anomalous long-range hard-wall repulsion, which cannot be categorized into any classical types. This repulsion is enhanced as either the affinity or the bulk cosolvent fraction increases, leading to a higher kinetic barrier to stabilize individual condensates. The emergence of the hard-wall repulsion is attributed to cosolvent regulation. The overlapping of the cosolvent excess layers triggers a local cosolvent condensation, dramatically increasing the energy.

Superior to the widely-used FH-based theories, our theory successfully captures the structural heterogeneity of the condensate, especially its cosolvent-rich interface, which is critical to regulate the interactions and stability. Our trapping approach enables us to directly track the energy and visualize the evolution of morphology as two condensates approach. Polymers in solvent mixtures is widely used to model biomolecular condensates in intracellular environment. The hard-wall repulsion predicted here demonstrates that cosolvent as an intrinsic ingredient in the system can itself regulate the interaction and stability of condensates. The existence and relative importance of other effects can only be evaluated when the essential cosolvent effect is rigorously considered.  

\begin{acknowledgments}
R. W. acknowledges the support from the University of California, Berkeley. This research used the computational resources provided by the Kenneth S. Pitzer Center for Theoretical Chemistry.
\end{acknowledgments}

\bibliographystyle{apsrev4-2}
\bibliography{Refs}

\end{document}


\textbf{\huge Supplementary Materials for}\\

\begin{center}
\vspace{0.4cm}
\textbf{\large Anomalous Long-range Hard-wall Repulsion between Polymers in Solvent Mixtures and Its Implication for Biomolecular Condensates}\\
\vspace{0.4cm}

\vspace{1\baselineskip}
Luofu Liu$^{1,2}$, and Rui Wang$^{1,2,*}$

\vspace{2\baselineskip}
1. Department of Chemical and Biomolecular Engineering, University of California Berkeley, CA 94720, USA

\vspace{1\baselineskip}
2. Materials Sciences Division, Lawrence Berkeley National Lab, Berkeley, California 94720, USA

\vspace{1\baselineskip}
* correspondence to: \textcolor{blue}{\underline {ruiwang325@berkeley.edu}}

\end{center}

\renewcommand*{\citenumfont}[1]{S#1}
\renewcommand*{\bibnumfmt}[1]{(S#1)}

\renewcommand{\theequation}{S\arabic{equation}}
\renewcommand{\thefigure}{S\arabic{figure}}


\section{I. Derivation of the Variational Approach to Control the Center-of-Mass of Polymers in a Solvent Mixture}

In this section, we provide a detailed derivation of the variational approach to control the center of mass (c.m.) of two polymers separately in a solvent mixture.

We consider a system of two identical polymer chains (P1 and P2) immersed in $n_{\rm S}$ solvent and $n_{\rm C}$ cosolvent molecules. The system is connected to a reservoir of binary mixture of solvents to maintain fixed chemical potentials $\mu_{\rm S}$ and $\mu_{\rm C}$ of solvent and cosolvent, respectively. The solvent is treated as a small molecule and the cosolvent is treated as an oligomeric chain. Both the polymer and the cosolvent are described by the continuous Gaussian-chain containing $N$ and $N_{\rm C}$ Kuhn segments, respectively, with Kuhn length $b$. The volumes of a Kuhn segment and a solvent are assumed to be the same $v$. To study the interaction between two polymers, we fix their c.m. at $\bm{\xi}_1$ and $\bm{\xi}_2$ respectively with the c.m. separation distance $L = |\bm{\xi}_2-\bm{\xi}_1|$. The semi-canonical partition function is
\begin{equation}
\begin{aligned}
    &Z = \sum_{n_{\rm S}, n_{\rm C} =0}^{\infty}\frac{e^{\beta\mu_{\rm S}n_{\rm S}} e^{\beta\mu_{\rm C}n_{\rm C}}}{n_{\rm S}!n_{\rm C}!v^{2N +n_{\rm S} +n_{\rm C}N_{\rm C}}} 
    \prod_{\alpha=1}^2 \int\hat{{\rm D}}\{\bm{R}_{\alpha}\} \prod_{\gamma=1}^{n_{\rm S}} \int {\rm d} {\bm{r}_\gamma} \prod_{\kappa=1}^{n_{\rm C}} \int\hat{{\rm D}}\{\bm{R}_{\kappa}\} \exp(-\beta H)\\
    &\prod_{\bm{r}} \delta \left[1 - \sum_{j={\rm P1}, {\rm P2}, {\rm S}, {\rm C}} \hat{\phi}_j(\bm{r}) \right] \prod_{\alpha=1}^2 \delta \left[\frac{1}{N}\int_0^N {\rm d}s \bm{R}_\alpha -\bm{\xi}_\alpha \right]
\end{aligned}
\label{eq:particle_partition_function}
\end{equation}
where $\int {\rm \hat{D}} \{\bm{R}_\alpha\}=\int {\rm D} \{\bm{R}_\alpha\} {\rm exp} [ {-(3/2b^2) \int^N_0 ds (\partial{\bm R_\alpha}}/{\partial s})^2 ]$ and  $\int {\rm \hat{D}} \{\bm{R}_\kappa\}=\int {\rm D} \{\bm{R}_\kappa\} {\rm exp} [ {-(3/2b^2) \int^{N_{\rm C}}_0 ds (\partial{\bm R_\kappa}}/{\partial s})^2 ]$ denote the functional integration over all possible chain configurations weighted by Gaussian statistics for polymer and cosolvent, respectively. $\int {\rm d} {\bm{r}_\gamma}$ integrates the degree of freedom of solvent molecules. The Hamiltonian
\begin{equation}\label{eq:hamiltonian}
    \beta H = v^{-1}\sum_{j,k}\chi_{jk}\int{\rm d}\bm{r}\hat{\phi}_j(\bm{r})\hat{\phi}_k(\bm{r}) \ \ \ (j,k={\rm P}, {\rm S}, {\rm C})
\end{equation}
includes interactions among the polymers, solvents, and cosolvents in terms of the Flory-Huggins $\chi$ parameters. The instataneous volume fractions are given by
\begin{subequations}
\begin{align}
    &\hat{\phi}_{\rm P\alpha}(\bm{r}) = v\int_0^N\delta(\bm{r} - \bm{R}_{\alpha}(s)) {\rm d}s \\
    & \hat{\phi}_{\rm S}(\bm{r}) = v\sum_{\gamma=1}^{n_{\rm S}}\delta(\bm{r} - \bm{r}_\gamma) \\
    &\hat{\phi}_{\rm C}(\bm{r}) = v\sum_{\kappa=1}^{n_{\rm C}}\int_0^{N_{\rm C}}\delta(\bm{r} - \bm{R}_{\kappa}(s)) {\rm d}s 
\end{align}
\end{subequations}
and $\hat{\phi}_{\rm P}(\bm{r}) = \hat{\phi}_{\rm P1}(\bm{r}) + \hat{\phi}_{\rm P2}(\bm{r})$ is the total instantaneous volume fraction of polymer. The first $\delta$-functional accounts for the incompressibility and the second enforces the c.m. constraint on the two polymers.

The transformation from the particle-based to the field-based representation is achieved by inserting the following identities into the partition function:
\begin{align}
1 =& \int {\rm D}\phi_{j} \prod_{\bm r}\delta [ \phi_{j}({\bm r})-\hat{\phi}_{j}({\bm r}) ]
=\int {\rm D}\phi_{j} {\rm D}W_{j} \exp \left\{ i \int d{\bm r} W_{j}({\bm r})
[ \phi_{j}({\bm r})-\hat{\phi}_{j}({\bm r}) ] \right\}  ~~(j={\rm P1, P2, S, C})
\end{align}
where the right-hand side of the equation arises from the Fourier representation of the $\delta$-function with $W_{j}({\bm r})$ being the Fourier conjugate field to
$\phi_{j}({\bm r})$.
Similarly, the Fourier representations of the incompressibility condition and constraint of the polymer c.m. are
\begin{align}
\delta \left[1 - \sum_{j} \hat{\phi}_j(\bm{r}) \right]
= \int {\rm D}\Gamma \exp \left\{ i \int {\rm d}{\bm r}\Gamma({\bm r})
\left[1 - \sum_{j} \hat{\phi}_j(\bm{r})\right] \right\}
\end{align}
\begin{align}
\delta \left[ \frac{1}{N}\int_0 ^N {\rm d}s {\bm R}_\alpha(s) - {\bm \xi}_\alpha \right] 
=
\int {\rm d} {\bm F}_\alpha  \exp \left\{ i {\bm F}_\alpha  \cdot
\left[ \frac{1}{N}\int_0 ^N {\rm d}s {\bm R}_\alpha (s) - {\bm \xi}_\alpha  \right]  \right\}
\end{align}
where $\Gamma$ is the field accounting for incompressibility and ${\bm F}_{\alpha}$ is the force field conjugate to the deviation of c.m. of the polymer $\alpha$ from the targeted position $\bm{\xi}_\alpha$.  By performing the identity transformations, we obtain the field-based partition function as
\begin{align}\label{SI_eq:field_based_partition_function}
&Z =
\prod_{j} \left(\int{\rm D}\phi_j\int{\rm D}W_j\right) {\rm D} \Gamma {\rm d}{\bm F}_1 {\rm d}{\bm F}_2
\sum_{n_{\rm S}, n_{\rm C} =0}^{\infty}\frac{e^{\beta\mu_{\rm S}n_{\rm S}} e^{\beta\mu_{\rm C}n_{\rm C}}}{n_{\rm S}!n_{\rm C}!v^{2N +n_{\rm S} +n_{\rm C}N_{\rm C}}} 
    \prod_{\alpha=1}^2 \int\hat{{\rm D}}\{\bm{R}_{\alpha}\} \prod_{\gamma=1}^{n_{\rm S}} \int {\rm d} {\bm{r}_\gamma} \prod_{\kappa=1}^{n_{\rm C}} \int\hat{{\rm D}}\{\bm{R}_{\kappa}\}
\nonumber \\
&\cdot \exp \left\{
\frac{1}{v} \int {\rm d}{\bm r} \left[
-\sum_{j,k}\chi_{jk}\phi_{j}(\bm{r})\phi_{k}(\bm{r})
+\sum_{j}iW_j(\bm{r}) \left( \phi_{j}({\bm r}) - \hat{\phi}_{j}({\bm r}) \right)
+i\Gamma ({\bm r}) \left( \sum_{j} \hat{\phi}_j(\bm{r})  - 1 \right) \right] \right\} \nonumber \\
&\cdot \exp \left\{ i {\bm F}_1 \cdot
\left[ \frac{1}{N}\int_0 ^N {\rm d}s {\bm R}_1(s) - {\bm \xi}_1 \right]  \right\} \exp \left\{ i {\bm F}_2 \cdot
\left[ \frac{1}{N}\int_0 ^N {\rm d}s {\bm R}_2(s) - {\bm \xi}_2 \right]  \right\}
\end{align}
where $j,k={\rm P1, P2, S, C}$. 

Here, Eq. \ref{SI_eq:field_based_partition_function} can be simplified and decomposed as $Z=Z_0 Z_{F1}Z_{F2}$, where $Z_0$ denotes the contribution irrelevant to the trap and also exists in other constraint-free systems. $Z_{F\alpha}$ ($\alpha=1,2$) comes from the constraint of the c.m. $Z_0$ is given by
\begin{equation}
\begin{aligned}
    &Z_0 = \prod_{j} \left(\int{\rm D}\phi_j\int{\rm D}W_j\right) {\rm D} \Gamma 
    \exp\bigg\{ e^{\beta\mu_{\rm S}}Q_{\rm S} + e^{\beta\mu_{\rm C}}Q_{\rm C} + \frac{1}{v}\int {\rm d}{\bm{r}}\bigg[-\sum_{j,k}\chi_{jk}\phi_j(\bm{r})\phi_k(\bm{r})  +\sum_{j}iW_{j}(\bm{r})\phi_j(\bm{r}) \\
     &+ i\Gamma(\bm{r})\left(\sum_{j} {\phi}_j(\bm{r})-1\right) \bigg] \bigg\}
\end{aligned}
\end{equation}
where the solvent partition function is given by $Q_{\rm S} = v^{-1}\int{\rm {d}}\bm{r}e^{-iW_{\rm S}(\bm{r})}$ and the cosolvent partition function $Q_{\rm C}$ can be evaluated by the chain propagator for cosolvent $q_{\rm C}(\bm{r},s)$ as $Q_{\rm C} = v^{-1}\int {\rm d}{\bm{r}}q_{\rm C}(\bm{r},s)q_{\rm C}(\bm{r},N_{\rm C}-s)$, with $q_{\rm C}(\bm{r},s)$ given by the modified diffusion equation 
\begin{equation}
   \frac{\partial q_{\rm C}}{\partial s} = \frac{b^2}{6}\nabla^2q_{\rm C}(\bm{r},s) - iW_{\rm C}(\bm{r})q_{\rm C}(\bm{r},s)
\end{equation}

$Z_{F\alpha}$ is given by
\begin{equation}
    Z_{F\alpha} = \int {\rm d} \bm{F}_\alpha \exp(-L_{{F}\alpha})
\end{equation}
where the ``action" $L_{F\alpha} = -\ln Q_{\rm \alpha}$, with $Q_{\alpha}$ the single-chain partition function of the polymer $\alpha$ in the auxiliary fields $W_{\rm P\alpha}$ and $\bm{F}_\alpha$:
\begin{equation}
\label{eq:single_chain_partition_function}
\begin{aligned}
    &Q_\alpha = \frac{1}{v^{N}}\int {\rm \hat{D}} \{{\bm{R}_\alpha}\} 
    \exp \bigg\{ -\int_0^N {\rm d}s \bigg[ iW_{{\rm P}\alpha}(\bm{R}_\alpha(s))
    - i\frac{\bm{F}_\alpha}{N} \cdot (\bm{R}_\alpha(s) - \bm{\xi}_\alpha)   \bigg]  \bigg\}
\end{aligned}
\end{equation}

To focus on the effect of the c.m. fluctuation, we perform a variational treatment on $Z_F$ while taking the saddle-point approximation to evaluate the functional integrals of fields included in $Z_0$ \cite{fredrickson2006}. In this way, the conjugate fields $W_j$ ($j={\rm P1, P2, S, C}$) and the incompressibility field $\Gamma$ will be replaced by their saddle-point values $-iw_{j}$ and $-i\gamma$, respectively. The free energy $\mathcal{W}_0$ corresponding to $Z_0$ is given by
\begin{equation}
\begin{aligned}
    &\mathcal{W}_0 = -\ln Z_0 \\
    &= -e^{\beta\mu_{\rm S}}Q_{\rm S} -e^{\beta\mu_{\rm C}}Q_{\rm C} + \frac{1}{v}\int {\rm d}{\bm{r}}\bigg[\sum_{j,k}\chi_{jk}\phi_j(\bm{r})\phi_k(\bm{r})  -\sum_{j}w_{j}(\bm{r})\phi_j(\bm{r})
     -\gamma(\bm{r})\left(\sum_{j} {\phi}_j(\bm{r})-1\right) \bigg] 
\end{aligned}
\end{equation}
To capture the c.m. fluctuation of the polymer which cannot be ignored when the polymer is in the swollen coil state, we use the Gibbs-Feynman-Bogoliubov variational approach \cite{kardar2007} to estimate the integral of $\bm F$ for evaluating $Z_F$:
\begin{equation}\label{eq:Feynman_bound}
    Z_{F\alpha} = Z_{\rm ref,\alpha} \langle \exp[-(L_{F\alpha}-L_{\rm ref,\alpha})] \rangle_{\rm ref}
    \approx Z_{\rm ref,\alpha} \exp[-\langle L_{F\alpha} - L_{\rm ref,\alpha} \rangle_{\rm ref}]
\end{equation}
where the average $\langle ... \rangle_{\rm ref}$ is taken in the reference ensemble with action $L_{\rm ref,\alpha}$.
$Z_{\rm ref,\alpha}=\int {\rm d} \bm{F}_\alpha \exp(-L_{\rm ref,\alpha})$ is the normalization factor. For mathematical convenience, $L_{\rm ref,\alpha}$ is taken to be a $2n$-power modified Gaussian
\begin{equation}
    e^{-L_{\rm ref,\alpha}} = \prod_{\kappa = x,y,z} (F_{\alpha, \kappa} + if_{\alpha,\kappa})^{2n} \exp \left[ - \frac{(F_{\alpha,\kappa} + if_{\alpha,\kappa})^2}{2A_{\alpha, \kappa}} \right]
\end{equation}
where the average force $\bm{f}_\alpha$ and coefficients $A_{\alpha,\kappa}$ are taken to be the variational parameters. The normalization factor $Z_{\rm ref,\alpha}$ can be calculated as
\begin{equation}
    Z_{\rm ref,\alpha}=\int {\rm d} \bm{F}_\alpha \exp(-L_{\rm ref,\alpha})=\prod_{\kappa=x,y,z}(2\pi)^{1/2}(2n-1)!!A^{n+1/2}_{\alpha,\kappa}
\end{equation}
This construction of $L_{\rm ref}$ assumes the independence of the c.m. fluctuations in the three directions. Note that a general $2n$-power modified Gaussian is needed instead of a standard ($0$th-power) Gaussian to reinforce the confinement of the c.m. as we demonstrated in our previous work \cite{Liu2024a}.

Evaluation of $e^{ - \langle L_{F\alpha} \rangle_{\rm ref} }$ gives an approximation of the single-chain partition function $Q_{\alpha}$ as follows:
\begin{equation}\label{eq:variational_SCPF}
Q_{\alpha} =
\frac{1}{v^{N}}\int \hat{{\rm D}} \{{\bm R}_\alpha\}
\exp \left\{
-\int_0^N {\rm d}s \ w_{\rm P\alpha}[{\bm R}_\alpha(s)]
+ {\bm f}_\alpha \cdot \overline{\Delta{\bm R}_\alpha} \right\}
\langle  \exp \left( i{\bm g}_\alpha \cdot \overline{\Delta{\bm R}_\alpha}  \right)  \rangle_{\rm ref}
\end{equation}
where ${\bm g}_\alpha={\bm F}_\alpha+i{\bm f}_\alpha$, and $\overline{\Delta{\bm R}_\alpha} = N^{-1} \int_0^N {\rm d}s [{\bm R}_\alpha(s) - {\bm \xi}_\alpha]$ is the difference of the instantaneous c.m. from the targeted position $\bm{\xi}_\alpha$. $\langle  \exp \left( i{\bm g}_\alpha \cdot \overline{\Delta{\bm R}_\alpha}  \right)  \rangle_{\rm ref}$ can be evaluated using the Gaussian integral as:
\begin{align}
\langle  \exp \left( i{\bm g}_\alpha \cdot \overline{\Delta{\bm R}_\alpha}  \right)  \rangle_{\rm ref} &=
\frac{1}{Z_{\rm ref,\alpha}} \prod_{\kappa=x,y,z}
\int {\rm d} g_{\alpha, \kappa} g^{2n}_{\alpha,\kappa}
\exp \left( i g_{\alpha,\kappa} \overline{\Delta R}_{\alpha,\kappa} - \frac{g^2_{\alpha,\kappa}}{2A_{\alpha,\kappa}}  \right)  \nonumber \\
&=\prod_{\kappa=x,y,z} \exp \left( - \frac{A_{\alpha,\kappa}\overline{\Delta R}^2_{\alpha,\kappa}}{2} \right)
\sum^n_{m=0} C_{n,m} (A_{\alpha,\kappa} \overline{\Delta R}^2_{\alpha,\kappa})^m
\end{align}
with the coefficient $C_{n,m}=(-1)^m[(2n)!(2n-2m-1)!!]/[(2m)!(2n-1)!!]$. The
single-chain partition function in Eq. \ref{eq:variational_SCPF} can then be written as
\begin{align}
Q_{\alpha} &=
\frac{1}{v^{N}}\int {\rm D} \{{\bm R}_\alpha\} e^{-H^*[\bm R_\alpha]}
\left[
\prod_{\kappa=x,y,z} \left( \sum^n_{m=0} C_{n,m} A^m_{\alpha,\kappa} \overline{\Delta R}^{2m}_{\alpha,\kappa}\right)
\right]
\end{align}
with the Hamiltonian $H^*[\bm{R}_\alpha]$ defined by
\begin{align}\label{eq:H_R}
H^*[{\bm R}_\alpha] &= \frac{3}{2b^2} \int_0^N {\rm d}s
 \left[ \frac{\partial{\bm R}_\alpha(s)}{\partial s} \right]^2
+\int_0^N {\rm d}s w_{\rm P\alpha}[{\bm R}_\alpha(s)]
- {\bm f}_\alpha \cdot \overline{\Delta{\bm R}_\alpha}
+ \sum_{\kappa=x,y,z} \frac{A_{\alpha,\kappa}}{2} \overline{\Delta R}^2_{\alpha,\kappa}
\end{align}
To facilitate the evaluation of $Q_\alpha$, we re-express it an ensemble average based on $H^*[\bm{R}]$:
\begin{align}
Q_{\alpha} &=
\frac{v^{-N}\int {\rm D} \{{\bm R}_\alpha\} e^{-H^*[\bm R_\alpha]}
\left[
\prod_{\kappa=x,y,z} \left( \sum^n_{m=0} C_{n,m} A^m_{\alpha,\kappa} \overline{\Delta R}^{2m}_{\alpha,\kappa}\right)
\right]}{v^{-N}\int {\rm D} \{{\bm R}_\alpha\} e^{-H^*[\bm R_\alpha]}}
\frac{1}{v^N} \int {\rm D} \{{\bm R}_\alpha\} e^{-H^*[\bm R_\alpha]} \nonumber \\
&=\left[ \prod_{\kappa=x,y,z} \left( \sum^n_{m=0} C_{n,m} A^m_{\alpha,\kappa}
\langle \overline{\Delta R}^{2m}_{\alpha,\kappa} \rangle_{Q^*_\alpha}
\right) \right] Q^*_\alpha
\end{align}
where $Q^*_\alpha$ is the single-chain partition function with the Hamiltonian $H^*[\bm R]_\alpha$
\begin{align}
Q^*_\alpha=\frac{1}{v^N} \int {\rm D} \{{\bm R}_\alpha\} e^{-H^*[\bm R_\alpha]}
\end{align}
A common way to evaluate $Q^*_\alpha$ requires replacing the complicated functional integral with the product of chain propagators. The chain propagators satisfy the modified diffusion equation \cite{fredrickson2006}. However, this becomes difficult here due to the nonlinear term appeared in $A_{\alpha,\kappa}\overline{\Delta R}^2_{\alpha,\kappa}$ in $H^*[\bm R]$ (Eq. \ref{eq:H_R}) which contains 2-body contributions.
To circumvent this difficulty, we decompose the Hamiltonian $H^*[\bm R]$ into a 1-body term $H^{(1)} [{\bm R}_\alpha(s)]$ and a 2-body term $H^{(2)} [{\bm R}_\alpha(s),{\bm R}_\alpha(t)]$:
\begin{align}\label{H1}
H^{(1)}[{\bm R}_\alpha(s)] &=  \int_0^N {\rm d}s
\left\{
\frac{3}{2b^2} \left[ \frac{\partial{\bm R}_\alpha(s)}{\partial s} \right]^2
+  w_{\rm P\alpha}[{\bm R}_\alpha(s)]
- \frac{\bm f_\alpha}{N} \cdot [{\bm R}_\alpha(s)-{\bm \xi_\alpha}]
+ \sum_{\kappa=x,y,z} \frac{A_{\alpha,\kappa}}{2} [R_{\alpha,\kappa}(s)-\xi_{\alpha,\kappa}]^2
\right\}
\end{align}
\begin{align}\label{H2}
H^{(2)}[{\bm R}_\alpha(s),{\bm R}_\alpha(t)] &=
\int_0^N {\rm d}s \int_0^N {\rm d}^{\prime}t
\left\{ \sum_{\kappa=x,y,z} \frac{A_{\alpha,\kappa}}{2} [R_{\alpha,\kappa}(s)-\xi_{\alpha,\kappa}][R_{\alpha,\kappa}(t)-\xi_{\alpha,\kappa}]
\right\}
\end{align}
where the prime in $\int^N_0 {\rm d}^{\prime}t$ indicates the case of $t=s$ is excluded in the integral. 
Here, for simplicity, we make a further assumption that the two-body intra-chain correlation term $H^{(2)}$ is negligible compared to the one-body term $H^{(1)}$ such that $H^* \approx H^{(1)}$. The whole derivation without this approximation can be found in our previous work \cite{Liu2024a}. Under this approximation, the last term in Eq. \ref{eq:H_R} can be evaluated by
\begin{equation}\label{eq:H_(1)}
\sum_{\kappa=x,y,z} \frac{A_{\alpha,\kappa}}{2} \overline{\Delta R}^2_{\alpha,\kappa} \approx \int_0^N {\rm d}s
 \sum_{\kappa=x,y,z} \frac{A_{\alpha,\kappa}}{2} [R_{\alpha,\kappa}(s)-\xi_{\alpha,\kappa}]^2
\end{equation}
This facilitates the re-expression of $Q^*_\alpha$ using the chain propagator $q(\bm{r},s)$ as $Q^*_\alpha = v^{-1} \int {\rm d} \bm{r} q_\alpha(\bm{r},N-s)q_\alpha(\bm{r},s)$. The chain propagator $q_\alpha$ satisfies:
\begin{equation}\label{eq:propagator}
    \frac{\partial q_\alpha}{\partial s} = \frac{b^2}{6}\nabla^2 q_\alpha(\bm{r} ,s)- U_{\rm P\alpha}(\bm{r}) q_\alpha(\bm{r} ,s)
\end{equation}
where $U_{\rm P\alpha}$ is the total field experienced by the polymer $\alpha$: 
\begin{equation}\label{eq:total_field_polymer}
U_{\rm P\alpha}(\bm{r}) = w_{\rm P\alpha}(\bm{r}) - \frac{1}{N} \bm{f}_\alpha\cdot (\bm {r} - \bm{\xi}_\alpha) + \sum_{\kappa=x,y,z} \frac{A_{\alpha,\kappa}}{2N} (\kappa -\xi_{\alpha,\kappa})^2
\end{equation}

Then the variational free energy corresponding to $Z_{F\alpha}$ is given by
\begin{equation}\label{eq:W_F}
\begin{aligned}
   &\mathcal{W}_{F\alpha} =-\ln Z_{F\alpha} = -\ln Z_{\rm ref,\alpha} -\langle L_{\rm ref,\alpha} \rangle_{\rm ref}+\langle L_{F\alpha}\rangle_{\rm ref} \\
   &= - \ln Q^*_\alpha
- \frac{1}{2} \sum_{\kappa=x,y,z} \ln A_{\alpha,\kappa}
- \sum_{\kappa=x,y,z} \ln \left( \sum^n_{m=0} C_{n,m} A^m_{\alpha,\kappa} \langle \overline{\Delta R}^{2m}_{\alpha,\kappa} \rangle_{Q^*_\alpha} \right) 
\end{aligned}
\end{equation}
where the term $- \frac{1}{2} \sum_{\kappa=x,y,z} \ln A_{\alpha,\kappa}$ comes from $-\ln Z_{\rm ref,\alpha}=-(n+1/2)\sum_{\kappa} \ln A_{\alpha,\kappa}$ and $-\langle L_{\rm ref,\alpha} \rangle_{\rm ref}=n\sum_{\kappa} \ln A_{\alpha,\kappa}$. Note that we ignore the constants independent of the variational parameters. The last term is from the $2n$-power in the modified Gaussian reference.
Using $\langle x^{2m} \rangle = \int {\rm d}x x^{2m} e^{-ax^2/2+bx} / \int {\rm d}x e^{-ax^2/2+bx}=\sum^m_l (-1)^l (2m-1)!! C_{m,l} b^{2l}a^{-(m+l)}$, the moments $\langle \overline{\Delta R}^{2m}_{\alpha,\kappa} \rangle_{Q^*}$ can be estimated by neglecting the elastic energy and interaction field $w_{\rm p}$ in Eq. \ref{eq:H_R}.
Thereafter, the last term in Eq. \ref{eq:W_F} becomes a logarithm of linear superposition of power functions for the variational parameter $A_{\alpha,\kappa}$, in the form of $-\sum_{\kappa} \ln ( \sum^n_{p=0} d_{\kappa,p} A^{-p}_{\alpha,\kappa} )$. $d_{\kappa,p}$ are prefactors independent of $A_{\alpha,\kappa}^{-p}$.

Minimizing $\mathcal{W}_{F\alpha}$ with respect to $A_{\alpha,\kappa}$, we obtain
\begin{align}\label{dW_dA=0}
-\frac{1}{Q_\alpha^*} \frac{\partial Q_\alpha^*}{\partial A_{\alpha,\kappa}}=
 \frac{1}{A_{\alpha,\kappa}}
\left[ \frac{1}{2}
+ \frac{\sum^n_{p=0} (-p) d_{\kappa,p} A^{-p}_{\alpha,\kappa}}{\sum^n_{p=0} d_{\kappa,p} A^{-p}_{\alpha,\kappa}}
\right] ~~(\kappa=x,y,z)
\end{align}
with the help of Eq. \ref{eq:H_R}, the left-hand side yields
\begin{equation}\label{Rg_square}
-\frac{1}{Q_\alpha^*} \frac{\partial Q_\alpha^*}{\partial A_{\alpha,\kappa}}=
\frac{1}{2}  \mathcal{R}^2_{\alpha,\kappa}  
\end{equation}
where $\mathcal{R}_{\alpha,\kappa}$ is the $\kappa$-component of the mean-square radius of gyration of polymer $\alpha$, given by
\begin{equation}
    \mathcal{R}^2_{\alpha,\kappa}= \frac{\int {\rm d}{\bm r} (\kappa-\xi_{\alpha,\kappa})^2 \phi_{\rm P\alpha}(\bm r) }{\int {\rm d}{\bm r} \phi_{\rm P\alpha}(\bm r)}
\end{equation}
For the right-hand side of Eq. \ref{dW_dA=0}, we neglect the $A_{\alpha,\kappa}$-dependence of the ratio $r=\sum^n_{p=0} (-p) d_{\kappa,p} A^{-p}_{\alpha,\kappa}/\sum^n_{p=0} d_{\kappa,p} A^{-p}_{\alpha,\kappa}$, and treat $r$ as a general fitting parameter for simplicity.
Then Eq. \ref{dW_dA=0} becomes
\begin{align}\label{dW_dA=0_final}
A_{\alpha,\kappa}=\frac{\lambda}{\mathcal{R}^2_{\alpha,\kappa}} ~~(\kappa=x,y,z)
\end{align}
with $\lambda=1+2r$.
Similarly, $\mathcal{W}_{F\alpha} \approx - \ln Q^*_{\alpha} - (\lambda/2) \sum_{\kappa} \ln A_{\alpha,\kappa}$.
Especially, if we take the standard (0-th power) Gaussian reference, we have $\lambda = 1$.
While the ratio $r$ arising from the $2n$-power modified Gaussian makes $\lambda$ adjustable, so that we can calibrate its value based on the criterion of the known size of the infinitely long ideal chain in a single solvent with $\chi=0.5$: $\mathcal{R}_\alpha^2 = Nb^2/6$ as $N \to \infty$. The detailed calibration process is shown in Ref \cite{Liu2024a}.

Taking the saddle-point approximation for densities $\phi_{j}(\bm{r})$, conjugate fields $w_{j}(\bm{r})$, incompressibility field $\gamma(\bm{r})$, and optimizing $\mathcal{W}_{F\alpha}$ with respect to the average force ${\bm f}_\alpha$ and coefficients $A_{\alpha,\kappa}$, we obtain
\begin{subequations}\label{eq:scf}
\begin{align}
    &w_{\rm P\alpha}(\bm{r})  = \chi_{\rm PS}\phi_{\rm S}(\bm{r})  + \chi_{\rm PC}\phi_{\rm C}(\bm{r}) +\gamma(\bm{r}) \\
    &w_{\rm S}(\bm{r})  = \chi_{\rm PS}\phi_{\rm P}(\bm{r})  + \chi_{\rm SC}\phi_{\rm C}(\bm{r}) +\gamma(\bm{r}) \\
    &w_{\rm C}(\bm{r})  = \chi_{\rm PC}\phi_{\rm P}(\bm{r})  + \chi_{\rm SC}\phi_{\rm S}(\bm{r}) +\gamma(\bm{r}) \\
    &\phi_{\rm P\alpha}(\bm{r}) = \frac{1}{Q_\alpha^*} \int {\rm d}s q_\alpha(\bm{r},N-s)q_\alpha(\bm{r},s) \\
    & \phi_{\rm S}(\bm{r}) = \exp[\beta \mu_{\rm s} - w_{\rm s}(\bm{r})]  \\
    &\phi_{\rm C}(\bm{r}) = e^{\beta\mu_{\rm C}}\int {\rm d}s q_{\rm C}(\bm{r},N-s)q_{\rm C}(\bm{r},s) \\
    &  \int {\rm d} \bm{r} (\bm{r} - \bm{\xi}_\alpha) \phi_{\rm P\alpha}(\bm{r}) = \bm{0}  \\
    & A_{\alpha, \kappa} = \lambda / \mathcal{R}_{\alpha, \kappa}^2 
\end{align}
\end{subequations}
as equations 6a-6h in the main text.
The resulting free energy of the system is
\begin{equation}\label{eq:free_energy}
\begin{aligned}
&\mathcal{W} = \mathcal{W}_0+ \sum_{\alpha=1}^{2} \mathcal{W}_{F\alpha} \\&= -e^{\beta\mu_{\rm S}}Q_{\rm S} -e^{\beta\mu_{\rm C}}Q_{\rm C} - \sum_{\alpha=1}^2\left(\ln Q_\alpha^* + \frac{\lambda}{2} \sum_{\kappa=x,y,z} \ln A_{\alpha,\kappa}\right)
\\ & +\frac{1}{v}\int {\rm d}{\bm{r}}\bigg[\sum_{j,k}\chi_{jk}\phi_j(\bm{r})\phi_k(\bm{r})  -\sum_{j}w_{j}(\bm{r})\phi_j(\bm{r})
     -\gamma(\bm{r})\left(\sum_{j} {\phi}_j(\bm{r})-1\right)\bigg]
\end{aligned}
\end{equation}
The potential of mean force $U$ can now be calculated as the free energy at given separation distance $L$ excess to two infinitely separated polymers: $U(L)=\mathcal{W}(L)-\mathcal{W}(\infty)$.

To find the relation between the chemical potentials of solvent $\mu_{\rm S}$ and cosolvent $\mu_{\rm C}$, we apply the above SCFT to the trivial homogeneous bulk solvent mixture. The free energy of the bulk is given by
\begin{equation}
    \beta \mathcal{W}^{\rm bulk}\cdot \frac{v}{V^{\rm bulk}} = \chi_{\rm SC}\phi_{\rm S}^{\rm bulk}\phi_{\rm C}^{\rm bulk} -\phi_{\rm S}^{\rm bulk} + \phi_{\rm S}^{\rm bulk}\ln\phi_{\rm S}^{\rm bulk} - \frac{\phi_{\rm C}^{\rm bulk}}{N_{\rm C}}  + \frac{\phi_{\rm C}^{\rm bulk}}{N_{\rm C}}\ln \phi_{\rm C}^{\rm bulk} - \beta \mu_{\rm S} \phi_{\rm S}^{\rm bulk} - \beta \mu_{\rm C}\frac{\phi_{\rm C}^{\rm bulk}}{N_{\rm C}}
\end{equation}
where $V^{\rm bulk}$ is the volume of the bulk solvent mixture. $\phi_{\rm S}^{\rm bulk}$ and $\phi_{\rm C}^{\rm bulk}$ are the solvent and cosolvent volume fractions in the bulk, satisfying $\phi_{\rm S}^{\rm bulk} + \phi_{\rm C}^{\rm bulk}=1$. The chemical potentials  $\mu_{\rm S}$ and $\mu_{\rm C}$ can be obtained by minimizing $\mathcal{W}^{\rm bulk}$ with respect to the numbers of solvent $n_{\rm S} = \frac{V^{\rm bulk}}{v}\phi_{\rm S}^{\rm bulk}$ and cosolvent $n_{\rm C} = \frac{V^{\rm bulk}}{v}\frac{\phi_{\rm C}^{\rm bulk}}{N_{\rm C}}$ as
\begin{equation}
    \frac{\partial \mathcal{W}^{\rm bulk}}{\partial n_{\rm S}} = \frac{\partial \mathcal{W}^{\rm bulk}}{\partial n_{\rm C}} = 0
\end{equation}
which leads to
\begin{equation}
\begin{aligned}
    &\beta\mu_{\rm S}=-1+\ln\phi_{\rm S}^{\rm bulk}+\left(1-\frac{1}{N_{\rm C}}\right)\phi_{\rm C}^{\rm bulk} + \chi_{\rm BC}(\phi_{\rm C}^{\rm bulk})^2 \\
    &\beta\mu_{\rm C}=-1+\ln\frac{\phi_{\rm C}^{\rm bulk}}{N_{\rm C}}+\left(1-{N_{\rm C}}\right)\phi_{\rm S}^{\rm bulk} + \chi_{\rm BC}N_{\rm C}(\phi_{\rm S}^{\rm bulk})^2     
\end{aligned}    
\end{equation}


\bibliographystyle{apsrev4-2}
\bibliography{Refs}